\documentclass[aps, pra, reprint
]{revtex4-1}
\usepackage{natbib}
\usepackage{graphicx}
\usepackage{xcolor}

\newcommand{\dt}[0]{\frac{\mathrm{d}}{\mathrm{d}t}}

\newcommand{\nt}[0]{\notag\\}
\newcommand{\hsp}[0]{\hspace{0.5cm}}
\newcommand{\ii}[0]{\mathrm{i}}
\usepackage{physics}
\begin{document}
\title{Non-Markovian Quantum Heat Statistics with the Reaction Coordinate Mapping}
\author{Mike Shubrook}
\email{mikejshubrook@gmail.com}
\affiliation{Quantum Engineering Centre for Doctoral Training, H. H. Wills Physics Laboratory and Department of Electrical and Electronic Engineering, University of Bristol, BS8 1FD, United Kingdom}
\affiliation{Department of Physics and Astronomy, University of Manchester, Oxford Road, Manchester M13 9PL, United Kingdom}
\author{Jake Iles-Smith}
\affiliation{School of Mathematical and Physical Sciences, University of Sheffield, Sheffield, S10 2TN, United Kingdom}
\affiliation{Department of Physics and Astronomy, University of Manchester, Oxford Road, Manchester M13 9PL, United Kingdom}
\author{Ahsan Nazir}
\email{ahsan.nazir@manchester.ac.uk}
\affiliation{Department of Physics and Astronomy, University of Manchester, Oxford Road, Manchester M13 9PL, United Kingdom}

\begin{abstract}
    The definition of heat in quantum mechanics is ambiguous. Complications arise in particular when the coupling between a quantum system and a thermal environment is non-negligible, as the boundary between the two becomes blurred, making the distinction between system and environment difficult to draw. The reaction coordinate mapping can be used in such regimes to redraw the boundary between the system and environment. In this paper we combine the reaction coordinate technique with a two-point measurement protocol to compare two different definitions of heat:~energetic changes with respect to the full environment Hamiltonian (prior to the mapping), and energetic changes with respect to the residual environment Hamiltonian (after the mapping). We find that the latter definition displays behaviour more expected of a heat bath in the highly non-Markovian regime considered.
    \end{abstract}

\maketitle

\section{Introduction} 
\label{sec: introduction}
In classical thermodynamics a distinction is drawn between two 
classes of observable quantities:
A \emph{state function} is a physical quantity, such as temperature or internal energy, that is well defined for each point in the system’s phase space. In contrast, a \emph{path function} is a physical quantity that depends on the specific path taken between two points in this phase space, for example heat and work. In a quantum mechanical setting, 
state functions can generally be represented as the trace of some Hermitian operator with the state of the system, or deduced from the state itself. However, path functions, such as work, do not have a clear analogue~\cite{talkner2007fluctuation}. 
Notably, this has generated significant debate within the quantum thermodynamics community regarding the 
appropriate definition of path functions like heat and work in quantum systems~\cite{esposito2009nonequilibrium, talkner2020colloquium, vinjanampathy2016quantum, silaev2014lindblad, binder2015operational, schmidt2015work, alhambra2018work, jarzynski2004nonequilibrium, mitchison2019quantum}.

In regimes where system-environment interactions are weak, and are therefore accurately captured by a Born-Markov master equation, heat can quite naturally be identified as the energy irreversibly emitted into (or absorbed from) the environment~\cite{talkner2020colloquium}.
However, when the interaction energy becomes comparable to the internal energy of the system, 
there is no longer a clear partition between system and environment degrees of freedom~\cite{barnett2002methods, fox2006quantum} as the two 
become strongly correlated, 
potentially exchanging energy and information in a non-Markovian (or reversible) fashion. 
This leads to further ambiguity as to how one should appropriately apportion the changes in internal and interaction energies into heat and work~\cite{jarzynski2004nonequilibrium, strasberg2016nonequilibrium, talkner2020colloquium, seifert2016first, schmidt2015work}. 

In this paper, we investigate the role that different system-environment partitions play on quantum heat statistics in the non-Markovian regime. To do so, we employ the reaction coordinate (RC) mapping of the spin-boson model~\cite{iles2014environmental}.
Here, a collective coordinate of the environment is incorporated into an enlarged effective system Hamiltonian (the \emph{extended system}), with the remaining environmental degrees of freedom included as a weakly-coupled \emph{residual environment}, which may be treated perturbatively using the reaction coordinate master equation (RCME). The resulting description has proven useful in studying the dynamics~\cite{iles2016energy,Maguire19frank, anto2021capturing} and thermodynamics~\cite{Newman17heat,  Strasberg2016reaction, nazir2018reaction, anto2021strong} of quantum systems in regimes of strong and non-Markovian system-environment interactions.   
We extend the reaction coordinate formalism to the two-point measurement protocol (TPMP)~\cite{esposito2009nonequilibrium,Restrepo18from,Restrepo19electron} to derive a \emph{heat-counting reaction coordinate master equation} (HC-RCME). This generalised master equation allows us to calculate the characteristic function that generates the stochastic heat probability distribution for strong system-environment interactions, which we also successfully benchmark in some exactly solvable cases. 

Central to this work are the two possible definitions of heat provided by the RC formalism. Heat may naturally be defined as changes in measurements of the full environment Hamiltonian prior to the mapping, or changes in measurements of the residual environment Hamiltonian after the mapping.
We find that these two possible definitions of heat demonstrate qualitative and quantitative differences in the first two moments of their probability distributions. We also find corresponding differences in the change in ergotropy~\cite{allahverdyan2004maximal,francica2020quantum} and von Neumann entropy of the original system and the extended system 
over the process considered. Together, 
these results suggest that for non-Markovian systems, defining heat as changes in the energy of the residual environment is more in line with the classical definition of heat in thermodynamics.

The paper is arranged as follows. In Section~\ref{sec:TPMP} we cover the TPMP, outlining how the characteristic function of heat transfer can be written as the trace of a generalised density operator. In Section~\ref{sec: rcm} we show how the RC mapping can be used in conjunction with the TPMP to probe quantum heat statistics in strong coupling and non-Markovian regimes. We do this by deriving the HC-RCME which describes the evolution of the generalised density operator. We also introduce the two possible definitions of heat stated above. In Section~\ref{sec: cf} we analyse the characteristic functions associated with the two definitions of heat, before looking at the first two moments of the corresponding probability distributions in Section~\ref{sec: moments}. Differences in the first moment of these two definitions of heat motivate studying the 
transfer of ordered energy between the original system and full environment, and between the extended system and residual environment, 
which we cover in Section~\ref{sec: ergotropy}. We summarise and conclude in Section~\ref{sec: conc}.

\section{Counting statistics in the strong coupling regime} 
\label{sec: heat statistics}
Path functions do not have a clear and unambiguous definition in quantum systems; notably there are no unique Hermitian operators, and thus observables, associated with them. In order to define the statistics of path functions we consider performing projective measurements. The two-point measurement protocol~\cite{talkner2020colloquium, esposito2009nonequilibrium, silaev2014lindblad} is a framework which can be used to calculate the full counting statistics of the path function we want to define. However, within the TPMP there is ambiguity in choosing what basis we perform these projective measurements onto. In the strong coupling and non-Markovian regime that we consider, choosing a particular basis to perform the projective measurements onto is not always straightforward.

\begin{figure}
    \centering
    \includegraphics[width=0.495\textwidth]{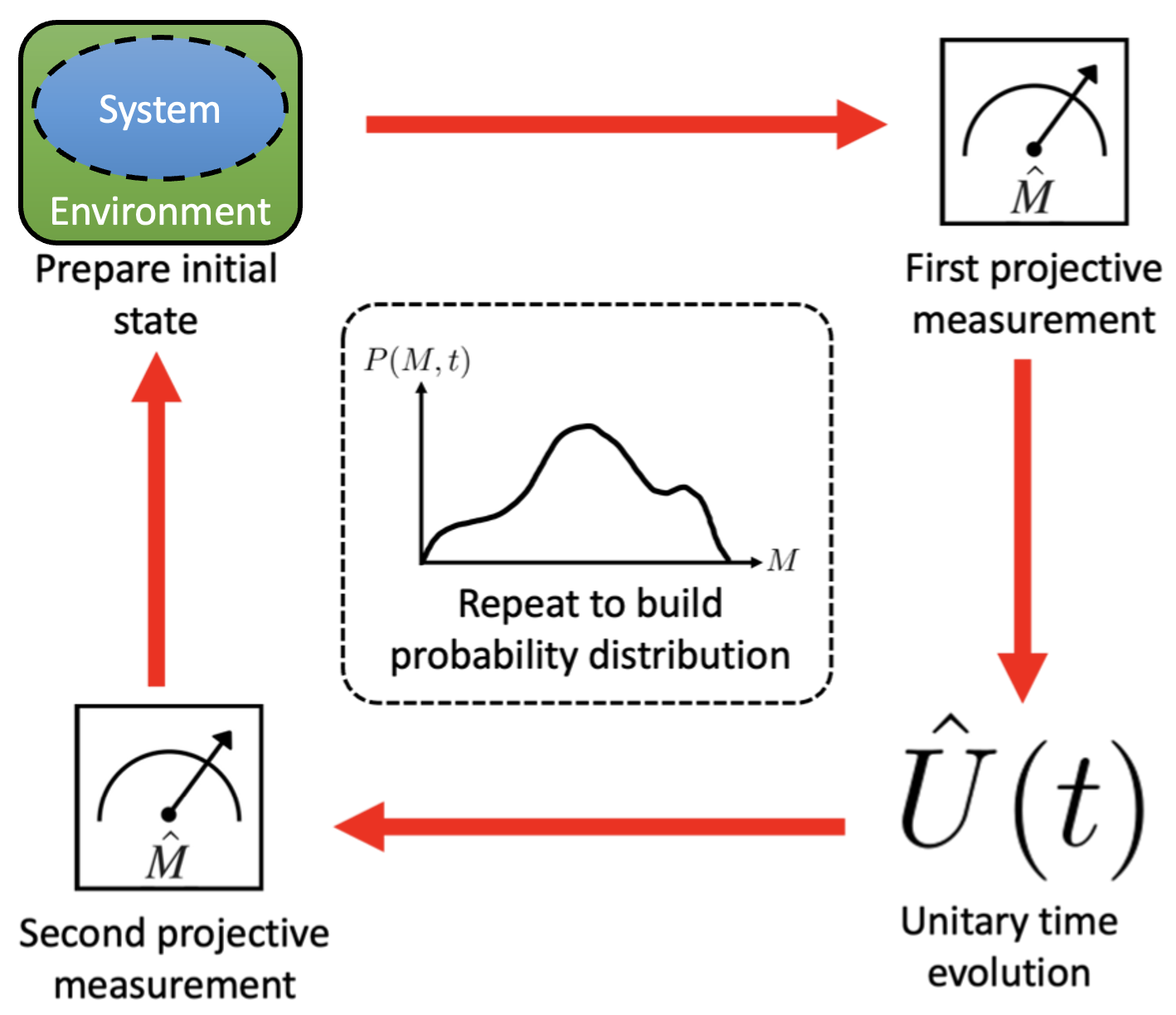}
    \caption{Schematic of the two-point measurement protocol. First, we prepare an initial state of the system, which we assumed couples to a thermal environment. We then perform a projective measurement onto the eigenbasis of some observable, $\hat{M}$, and allow the composite state to evolve unitarily for a time $t$, before performing another projective measurement onto the eigenbasis of $\hat{M}$. We take the difference of the two measurement outcomes, $M$, to be the value of the quantity we want to measure (in this case, heat). This value is stochastic, so we repeat the protocol many times in order to build a probability distribution, $P(M,t)$. 
}
    \label{fig:TPMP}
\end{figure}

\subsection{The two-point measurement protocol}
\label{sec:TPMP}
We define changes in heat through the TPMP, a brief overview of which can be seen schematically in Fig.~\ref{fig:TPMP} and is as follows: 
An open quantum system $S$ is coupled to an environment $E$, which together are governed by the Hamiltonian $\hat{H}$, evolving unitarily according to $\hat{U}(t) = \mathrm{exp}[-\ii \hat{H} t]$. 
We first prepare a product state $\hat{\rho}(0) = \hat{\rho}_{\mathrm{S}}(0)\otimes\hat{\rho}_{\mathrm{FE}}$, where $\hat{\rho}_{\mathrm{FE}}$ is a Gibbs state of the (full) environment and $\hat{\rho}_\mathrm{S}(0)$ is an arbitrary initial state of the system. We use the term `full environment' to describe the environment before we perform the reaction coordinate mapping. Next, a projective measurement onto the eigenbasis of an observable $\hat{M}$ is applied to the composite system, which is then allowed to evolve unitarily for time $t$, before performing a second projective measurement onto the eigenbasis of $\hat{M}$. 
We define the difference of these two measurement outcomes as $M$, a stochastic quantity. By repeating the protocol many times we build up a probability distribution $P(M,t)$. Our choice of operator $\hat{M}$ leads to the definition of heat that we use. 

While the probability distribution $P(M,t)$ contains the statistics that we are interested in, it is more convenient to work with its Fourier transform, the characteristic function (CF),
\begin{align}
\label{eq: CF PD}
    \Phi(\chi, t) &= \frac{1}{2\pi}\int\limits_{-\infty}^{\infty}\dd M P(M,t)\mathrm{e}^{\mathrm{i} M \chi},
\end{align}
a complex valued function of the counting parameter $\chi$, which is the conjugate parameter to the stochastic quantity $M$. 
Following Esposito \emph{et al.}~\cite{esposito2009nonequilibrium}, we can write the CF as
\begin{align}
\label{eq: CF 2}
    \Phi(\chi, t) &= \Tr[\hat{\rho}(\chi,t)],
\end{align}
where we have defined the generalised density operator 
\begin{align}
    \label{CFD density op}
    \hat{\rho}(\chi,t) = \hat{U}_{\frac{\chi}{2}}(t) \mathrm{e}^{-\ii\frac{\chi}{2} \hat{M}}\Bar{\rho}_{0} \mathrm{e}^{\ii\frac{\chi}{2} \hat{M}}\hat{U}_{-\frac{\chi}{2}}^{\dagger}(t),
\end{align}
and have made use of a generalised time evolution operator
\begin{align}
    \label{CFD time evolution op}
    \hat{U}_{\chi}(t) = \mathrm{e}^{\ii\chi \hat{M}}\hat{U}(t)\mathrm{e}^{-\ii\chi \hat{M}}.
\end{align}
In Eq.~\eqref{CFD density op}, $\Bar{\rho}_{0} = \sum_{m_{0}}\dyad{m_{0}}{m_{0}}\hat{\rho}(0)\dyad{m_{0}}{m_{0}}$ is the initial state of the composite system after being averaged over the initial measurement outcomes of $\hat{M}=\sum_{m_{0}}m_{0}\dyad{m_{0}}{m_{0}}$.
Applying a partial trace over the full environment degrees of freedom in Eq.~\eqref{eq: CF 2} gives
\begin{align}
\label{CF reduced}
    \Phi(\chi, t) &
    = \Tr_{\mathrm{S}}[\hat{\rho}_{\mathrm{S}}(\chi,t)],
\end{align}
where we have defined the reduced generalised density operator for the system as $\hat{\rho}_{\mathrm{S}}(\chi,t) = \Tr_{\mathrm{FE}}[\hat{\rho}(\chi,t)]$. By deriving an equation of motion for $\hat{\rho}_{\mathrm{S}}(\chi,t)$ and taking the trace, we can generate the full probability distribution for the stochastic quantity $M$ by inverting the Fourier transform.  

As shown in Ref.~\cite{silaev2014lindblad}, in the limit of weak system-environment interactions, 
we are able to derive a Born-Markov master equation describing the evolution of the reduced generalised density operator,
\begin{align}
\label{eq: weak coupling CFD ME}
    \dt \hat{\rho}_{\mathrm{S}}(\chi, t) =  \mathcal{L}_{\mathrm{S}}(\chi)\hat{\rho}_{\mathrm{S}}(\chi,t),
\end{align}
where $\mathcal{L}_{\mathrm{S}}(\chi)$ is the generalised ($\chi$-dressed) Liouville superoperator~\cite{breuer2002theory} that describes both the coherent dynamics and the effect of the environment on $\hat{\rho}_{\mathrm{S}}(\chi,t)$ up to second order in the coupling strength.
Solving this master equation with an appropriate initial condition, $\hat{\rho}_{\mathrm{S}}(\chi, 0)$, yields the CF.

\subsection{Reaction coordinate mapping} 
\label{sec: rcm}
In this paper we are interested in studying heat exchange beyond the Born-Markov limit, and therefore must move past the generalised Born-Markov equation given in Eq.~\eqref{eq: weak coupling CFD ME}. We consider the spin-boson model, 
\begin{align}
\label{eq: SBM Hamiltonian}
    \hat{H} &= \frac{\epsilon}{2}\hat{\sigma}_{z} + \frac{\Delta}{2}\hat{\sigma}_{x} + \hat{\sigma}_{z}\otimes\sum_{k}f_{k}(\hat{c}_{k}^{\dagger}+\hat{c}_{k}) + \sum_{k}\nu_{k}\hat{c}_{k}^{\dagger}\hat{c}_{k},
\end{align}
where the first two terms make up the system Hamiltonian $\hat{H}_{\mathrm{S}}$, the third term is the interaction $\hat{V}$, and the fourth is the self Hamiltonian of the full environment (FE), $\hat{H}_{\mathrm{FE}}$. In Eq.~\eqref{eq: SBM Hamiltonian}, $\epsilon$ is the energy gap between basis states $\ket{e}$ and $\ket{g}$ of a two-level system (TLS), and $\Delta$ is the tunneling between these states. The $k^{\mathrm{th}}$-mode of the bosonic environment has creation (annihilation) operators $\hat{c}_{k}^{\dagger}$ ($\hat{c}_{k}$), with energy $\nu_k$, and couples to the TLS with strength $f_{k}$. The TLS-full environment coupling can be described by the spectral density function, $J_{\mathrm{FE}}(\nu) = \sum_{k}\vert f_{k}\vert^2\delta(\nu - \nu_{k})$.

There has been significant effort in developing numerical methods to describe both the dynamics and thermodynamics of the spin-boson model. 
Notable examples include those based on discrete-time path integrals~\cite{makri1995tensorI,makri1995tensorII} and their expression as tensor networks~\cite{strathearn2018efficient,jorgensen2019exploiting}, 
hierarchical equations of motion~\cite{tanimura89heom, Ishizaki05heom,tang2015extended}, 
and chain mapping methods~\cite{10.1063/1.3490188,rosenbach2016efficient}.
While such techniques can be used to obtain numerically exact results in many regimes, they are computationally demanding and can be challenging to interpret. In contrast, nonperturbative master equation techniques, such as the polaron theory~\cite{mccutcheon2010quantum} and its variational extensions~\cite{10.1063/1.3636081, mccutcheon2011general} are intuitive and computational cheap, though often at the expense of restricted applicability. The nonperturbative method we use in this work is the reaction coordinate mapping (RCM)~\cite{iles2014environmental,strasberg2016nonequilibrium,Newman17heat,nazir2018reaction}.


To perform this mapping, a collective coordinate (the RC) is first extracted from the full environment. The TLS, RC and their interaction are then incorporated into an extended system (ES), and the remaining environment degrees of freedom are collected into the residual environment (RE), which couples to the extended system through the RC. For a more detailed discussion of the RC mapping we refer the reader to Refs.~\cite{iles2014environmental,nazir2018reaction}. 

Upon applying the RC mapping to the Hamiltonian in Eq.~\eqref{eq: SBM Hamiltonian} with the unitary operator $\hat{\mathcal{R}}$, we obtain the Hamiltonian,
\begin{align}
\label{eq: RC Hamiltonian}
    &\hat{H}_{\mathcal{R}}= \hat{\mathcal{R}}\hat{H}\hat{\mathcal{R}}^{\dagger} = \hat{H}_\mathrm{ES} + \hat{H}_\mathrm{I} + \hat{H}_\mathrm{RE},
    \\
    \label{eq:ES_HAM}
    &\hat{H}_\mathrm{ES} = \frac{\epsilon}{2}\hat{\sigma}_{z} + \frac{\Delta}{2}\hat{\sigma}_{x} + \Omega\hat{a}^{\dagger}\hat{a} + \lambda\hat{\sigma}_{z}(\hat{a}^{\dagger} + \hat{a}),\\
    \label{eq:RC_INT}
    &\hat{H}_\mathrm{I} = (\hat{a}^{\dagger} + \hat{a})\sum_{k}g_{k}(\hat{b}_{k}^{\dagger}+\hat{b}_{k}) 
    + (\hat{a}^{\dagger} + \hat{a})^{2}\sum_{k}\frac{g_{k}^{2}}{\omega_{k}},\\
    \label{eq:RE_ham}
    &\hat{H}_\mathrm{RE} = \sum_{k}\omega_{k}\hat{b}_{k}^{\dagger}\hat{b}_{k}.
\end{align}
Here, Eq.~\eqref{eq:ES_HAM} is the extended system Hamiltonian, 
where the RC has frequency $\Omega$, and creation (annihilation) operators $\hat{a}^{\dagger}$ ($\hat{a}$), 
and is coupled to the TLS with strength $\lambda$. The self Hamiltonian of the RC is given by $\hat{H}_{\mathrm{RC}}=\Omega\hat{a}^{\dagger}\hat{a}$.
Eq.~\eqref{eq:RC_INT} defines the interaction between the RC and the residual environment with a coupling strength $g_{k}$ between the RC and $k^{\mathrm{th}}$ mode. The interaction also includes a counter term which ensures that the Hamiltonian is bounded from below. Eq.~\eqref{eq:RE_ham} gives the residual environment Hamiltonian with modes of frequency $\omega_{k}$, and creation (annihilation) operators $\hat{b}_{k}^{\dagger}$ ($\hat{b}_{k}$). 

We choose an underdamped Drude-Lorentz spectral density for the full environment, describing a peak of width $\Gamma$, centered at $\omega_{0}$, with coupling strength $\alpha$,
\begin{align}
    \label{DL Jw}
    J_{\mathrm{FE}}(\omega) = \frac{\alpha\Gamma\omega_{0}^{2}\omega}{(\omega_{0}^{2}-\omega^{2})^{2} + (\Gamma\omega)^{2}}.
\end{align}
Upon applying the RC mapping, the spectral density describing the extended system-residual environment interaction becomes Ohmic,
\begin{align}
    \label{RC Jw}
    J_{\mathrm{RE}}(\omega) = \sum_{k}\vert g_{k}\vert^2\delta(\omega - \omega_{k}) = \gamma\omega,
\end{align}
with the coupling strength between the extended system and residual environment being $\gamma = {\Gamma}/{\omega_{0}}$. A hard cutoff of $\omega_{\mathrm{cut}}  = 10\omega_{0}$, is added to both spectral densities in later numerical simulations. This choice of cutoff ensures convergence of all the results that we show.

For sufficiently weak coupling to the residual environment (small $\gamma$) we can derive a Born-Markov master equation for the extended system state $\hat{\rho}_{\mathrm{ES}}(t)$. After solving this reaction coordinate master equation, we can trace out the RC degrees of freedom to obtain the reduced state of the TLS: $\hat{\rho}_{\mathrm{S}}(t) = \Tr_{\mathrm{RC}}[\hat{\rho}_{\mathrm{ES}}(t)]$. The RCME accurately captures both non-Markovian and strong coupling effects between the system and full environment, since $\gamma$ is independent of $\alpha$. To confirm this, we present a comparison of an exactly solvable limit of the spin-boson model (the independent boson model for which $\Delta=0$~\cite{breuer2002theory}) with the RCME in Appendix~\ref{app:bench}, demonstrating excellent agreement across all timescales tested within the non-Markovian regime. 

\subsection{Heat counting in the reaction coordinate formalism}
We now apply the RC mapping to heat counting statistics. Let us begin with the CF to count energetic changes in the full environment, given by 
\begin{align}
\label{eq: app: CF FB}
    \Phi_{\mathrm{F}}^{\mathrm{rc}}(\chi, t) = \Tr[\mathrm{e}^{\ii\chi \hat{H}_{\mathrm{FE}}}\hat{U}(t)\mathrm{e}^{-\ii\chi \hat{H}_{\mathrm{FE}}}\Bar{\rho}_{0}\hat{U}^{\dagger}(t)],
\end{align}
where we assume projective measurements onto the full environment Hamiltonian $\hat{M} = \hat{H}_{\mathrm{FE}}$~\footnote{These projective measurements are not intended to be carried out physically; rather, they are a mathematical tool used to derive the heat-counting reaction coordinate master equation.}. We make the assumption that the initial state is uncorrelated, $\hat{\rho}(0) = \hat{\rho}_{\mathrm{S}}(0)\otimes \hat{\rho}_{\mathrm{FE}}$, with $\hat{\rho}_{\mathrm{FE}}$ the Gibbs state $\hat{\rho}_{\mathrm{FE}} = \mathrm{exp}[-\beta \hat{H}_{\mathrm{FE}}]/\mathcal{Z}_{\mathrm{FE}}$, where $\mathcal{Z}_{\mathrm{FE}} = \mathrm{Tr}[\mathrm{exp}[-\beta \hat{H}_{\mathrm{FE}}]]$ is the partition function and $\beta$ is the inverse temperature of the full environment. 
By resolving the identity as $\hat{\mathcal{I}} = \hat{\mathcal{R}}^{\dagger} \hat{\mathcal{R}}$ we can rewrite the CF in Eq.~\eqref{eq: app: CF FB} as
\begin{align}
\label{eq: app: CF FB RCM}
    \Phi_{\mathrm{F}}^{\mathrm{rc}}(\chi, t) &= \mathrm{Tr}\bigg[\mathrm{e}^{\ii\chi \hat{\mathcal{R}}^{\dagger}\hat{H}_{\mathrm{FE}}\hat{\mathcal{R}}}\hat{U}_{\hat{\mathcal{R}}}(t)\mathrm{e}^{-\ii\chi \hat{\mathcal{R}}^{\dagger}\hat{H}_{\mathrm{FE}}\hat{\mathcal{R}}}\hat{\mathcal{R}}^{\dagger}\nt&\hsp\hsp\times\hat{\rho}(0)\hat{\mathcal{R}}\hat{U}_{\hat{\mathcal{R}}}^{\dagger}(t)\bigg].
\end{align}
We have also defined the time evolution operator in the reaction coordinate frame as 
\begin{align}
    \hat{U}_{\mathcal{R}}(t) =\hat{\mathcal{R}}\mathrm{exp}[-\ii\hat{H}t]\hat{\mathcal{R}}^{\dagger}=  \mathrm{exp}[-\ii\hat{H}_{\mathcal{R}}t],
\end{align} 
where $\hat{H}_{\mathcal{R}}$ is the mapped Hamiltonian given in Eq.~\eqref{eq: RC Hamiltonian}. 
We assume that the RC mapping is performed such that the interaction between the RC and residual environment is weak. Performing the RC mapping and making the weak coupling approximation gives us
\begin{align}
    \Phi_{\mathrm{F}}^{\mathrm{rc}}(\chi, t) &\approx \mathrm{Tr}\bigg[\mathrm{e}^{\ii\chi (\hat{H}_{\mathrm{RC}} + \hat{H}_{\mathrm{RE}})}\hat{U}_{\hat{\mathcal{R}}}(t)\mathrm{e}^{-\ii\chi (\hat{H}_{\mathrm{RC}} + \hat{H}_{\mathrm{RE}})}\nt&\hsp\hsp\times\hat{\rho}_{\mathrm{S}}(0)\otimes\hat{\rho}_{\mathrm{RC}}\otimes \hat{\rho}_{\mathrm{RE}} \hat{U}_{\hat{\mathcal{R}}}^{\dagger}(t)\bigg],
\end{align}
valid provided that the product $\gamma\chi$ does not become too large.
Here we have approximated the action of the RCM on the full environment thermal state as
\begin{align}
    \hat{\mathcal{R}}^{\dagger} \hat{\rho}_{\mathrm{FE}} \hat{\mathcal{R}} \approx \hat{\rho}_{\mathrm{RC}}\otimes \hat{\rho}_{\mathrm{RE}},
\end{align} 
where $\hat{\rho}_{\mathrm{RC}}$ and $\hat{\rho}_{\mathrm{RE}}$ are Gibbs states of the RC and residual environment, respectively.
Using the cyclic property of the trace we rewrite the CF as
\begin{align}
    \Phi_{\mathrm{F}}^{\mathrm{rc}} &= \mathrm{Tr}\big[\mathrm{e}^{\ii\chi \hat{H}_{\mathrm{RC}} }\hat{\rho}(\chi, t)\big],
\end{align}
where we have defined the generalised density operator as
\begin{align}
\label{eq: app: generalised state}
    \hat{\rho}(\chi, t) &= \mathrm{e}^{\ii\frac{\chi}{2} \hat{H}_{\mathrm{RE}} }\hat{U}_{\hat{\mathcal{R}}}(t)\mathrm{e}^{-\ii\chi (\hat{H}_{\mathrm{RC}} + \hat{H}_{\mathrm{RE}})}\nt&\hsp\times\hat{\rho}_{\mathrm{S}}(0)\otimes\hat{\rho}_{\mathrm{RC}}\otimes \hat{\rho}_{\mathrm{RE}}\hat{U}_{\hat{\mathcal{R}}}^{\dagger}(t)\mathrm{e}^{\ii\frac{\chi}{2} \hat{H}_{\mathrm{RE}} }.
\end{align}
By taking the time derivative of this generalised density operator and moving into the interaction picture with respect to $\hat{H}_{\mathrm{ES}}+\hat{H}_{\mathrm{RE}}$ we find an equation of motion which resembles the Liouville-von Neumann equation~\cite{breuer2002theory}
\begin{align}
    \dt \tilde{\rho}(\chi, t) = -\ii\bigg(\tilde{H}_{\mathrm{I}}(\chi, t)\tilde{\rho}(\chi, t) - \tilde{\rho}(\chi, t)\tilde{H}_{\mathrm{I}}(-\chi, t)\bigg),
\end{align}
where 
\begin{align}
    \tilde{H}_{\mathrm{I}}(\chi, t) = \mathrm{e}^{\ii\frac{\chi}{2} \hat{H}_{\mathrm{RE}} }  \mathrm{e}^{\ii (\hat{H}_{\mathrm{ES}} + \hat{H}_{\mathrm{RE}}) t} \hat{H}_{\mathrm{I}} \mathrm{e}^{-\ii (\hat{H}_{\mathrm{ES}} + \hat{H}_{\mathrm{RE}}) t} \mathrm{e}^{-\ii\frac{\chi}{2} \hat{H}_{\mathrm{RE}} }.
\end{align}
We use this equation as a basis to derive a master equation which treats the Hamiltonian of the extended system exactly and the effect of the residual environment on the extended system up to second order in $\hat{H}_{\mathrm{I}}$, leading us to the 
HC-RCME
\begin{align}
\label{eq: HC-RCME}
    \dt \hat{\rho}_{\mathrm{ES}}(\chi, t) &= \mathcal{L}_{\mathrm{ES}}(\chi)[\hat{\rho}_{\mathrm{ES}}(\chi, t)].
\end{align}
Details of the form of $\mathcal{L}_{\mathrm{ES}}(\chi)$ can be found in Appendix~\ref{app: HC-RCME}. By taking the partial trace over the residual environment degrees of freedom, the CF to count energetic changes in the full environment Hamiltonian is given by
\begin{align}
\label{eq: CF full RC}
    \Phi_{\mathrm{F}}^{\mathrm{rc}} &= \mathrm{Tr}_{\mathrm{ES}}\big[\mathrm{e}^{\ii\chi \hat{H}_{\mathrm{RC}} }\hat{\rho}_{\mathrm{ES}}(\chi, t)\big],
\end{align}
where $\hat{\rho}_{\mathrm{ES}}(\chi, t)$ is found by solving the HC-RCME.

Notice, however, that the RC mapping gives us two possible environments to use within the TPMP. There is the environment prior to the mapping, which includes all bosonic degrees of freedom (the full environment), and there is the environment after the mapping (the residual environment). 
Performing the projective measurements onto these two Hamiltonians results in two different CFs and thus definitions of heat, which we can compare.
When counting only on the residual environment we take $\hat{M} = \hat{\mathcal{R}}^{\dagger}\hat{H}_{\mathrm{RE}}\hat{\mathcal{R}}$, 
the residual environment Hamiltonian in the unmapped frame, which leads to 
\begin{align}
\label{eq: RC CF RB}
    \Phi_{\mathrm{R}}^{\mathrm{rc}}(\chi,t) &= \mathrm{Tr}_{\mathrm{ES}}\big[\hat{\sigma}_{\mathrm{ES}}(\chi,t)\big],
\end{align}
where the generalised density operator is given by
\begin{align}
    \hat{\sigma}(\chi, t) &= \mathrm{e}^{\ii\frac{\chi}{2} \hat{H}_{\mathrm{RE}} }\hat{U}_{\hat{\mathcal{R}}}(t)\mathrm{e}^{-\ii\chi \hat{H}_{\mathrm{RE}}}\nt&\hsp\times\hat{\rho}_{\mathrm{S}}(0)\otimes\hat{\rho}_{\mathrm{RC}}\otimes \hat{\rho}_{\mathrm{RE}}\hat{U}_{\hat{\mathcal{R}}}^{\dagger}(t)\mathrm{e}^{\ii\frac{\chi}{2} \hat{H}_{\mathrm{RE}} }.
\end{align}
The two generalised density operators $\hat{\rho}_{\mathrm{ES}}(\chi,t)$ and $\hat{\sigma}_{\mathrm{ES}}(\chi,t)$ both obey the HC-RCME but have different initial conditions, being
\begin{align}
\label{eq: rho chi 0}
    \hat{\rho}_{\mathrm{ES}}(\chi,0) &=\hat{\rho}_{\mathrm{S}}(0)\otimes\frac{\mathrm{e}^{-(\beta+\ii\chi)\hat{H}_{\mathrm{RC}}}}{\Tr[\mathrm{e}^{-\beta \hat{H}_{\mathrm{RC}}}]},
\end{align}
and 
\begin{align}
\label{eq: sigma chi0}
    \hat{\sigma}_{\mathrm{ES}}(\chi,0) &= \hat{\rho}_{\mathrm{S}}(0)\otimes\hat{\rho}_{\mathrm{RC}}.
\end{align}
Here, $\hat{\rho}_\mathrm{S}(0)$ is the initial state of the TLS, which we take to be $\hat{\rho}_{\mathrm{S}}(0) = \dyad{+}$ throughout, where $\ket{+} = \frac{1}{\sqrt{2}}(\ket{e}+\ket{g})$.

While the HC-RCME can in principle be used to investigate the full spin-boson model, in order to compare the difference in system-environment partitions on the resulting heat statistics it is instructive to consider the 
($\Delta=0$) independent boson model (IBM), which is exactly solvable since the system and interaction Hamiltonians now commute~{\footnote{{We note that the following results do not change qualitatively when a small driving term $\Delta>0$ is added.}}}.
Following Popovic \emph{et al.}~\cite{popovic2021quantum}, 
we we are able to find an exact analytic expression for the CF describing energetic changes in the full environment, given by
\begin{align}
    \label{eq: exact CF}
    \Phi_{\mathrm{F}}^{\mathrm{ex}}(\chi,t) &=\exp\bigg[-2\int\limits_{0}^{\infty}\dd\omega\frac{J_{\mathrm{FE}}(\omega)}{\omega^{2}}\big(1-\cos(\omega t)\big)\\ \notag &\times\bigg(\coth(\frac{\beta\omega}{2})\big(1-\cos(\omega\chi)\big)-\ii\sin(\omega\chi)\bigg)\bigg].
\end{align}
A similar analytic expression for changes in the residual environment is not available since this definition relies on performing the RC mapping, with the result that the mapped system and interaction Hamiltonians no longer commute.
Despite being a dissipation-free model, the IBM is valuable for studying heat statistics as it enables benchmarking of the HC-RCME, while highlighting non-Markovian and strong-coupling effects. There has also been recent interest in the thermodynamics of pure decoherence processes, see e.g. Ref.~\cite{popovic2023thermodynamics}.

\section{Results}
The results presented here are based on three distinct CFs. Two of these CFs quantify energetic changes in the full environment Hamiltonian, using either the approximate (Eq.~\eqref{eq: CF full RC}) or exact (Eq.~\eqref{eq: exact CF}) methods of calculating the dynamics of the generalised density operator (i.e. by solving the HC-RCME or the IBM dynamics, respectively). The third CF (Eq.~\eqref{eq: RC CF RB}) quantifies energetic changes in the residual environment Hamiltonian  and is obtained by solving the HC-RCME with the appropriate initial condition. We use subscripts to denote the definition of heat associated with the CF: `F' for energetic changes in the full environment and `R' for energetic changes in the residual environment. The superscript denotes the method used for calculating the generalized density operator dynamics: `rc' for the HC-RCME and `ex' for the analytical dynamics derived from the IBM.

\subsection{Characteristic function}
\label{sec: cf}
\begin{figure*}[t]
     \centering
        \includegraphics[width=0.995\textwidth]{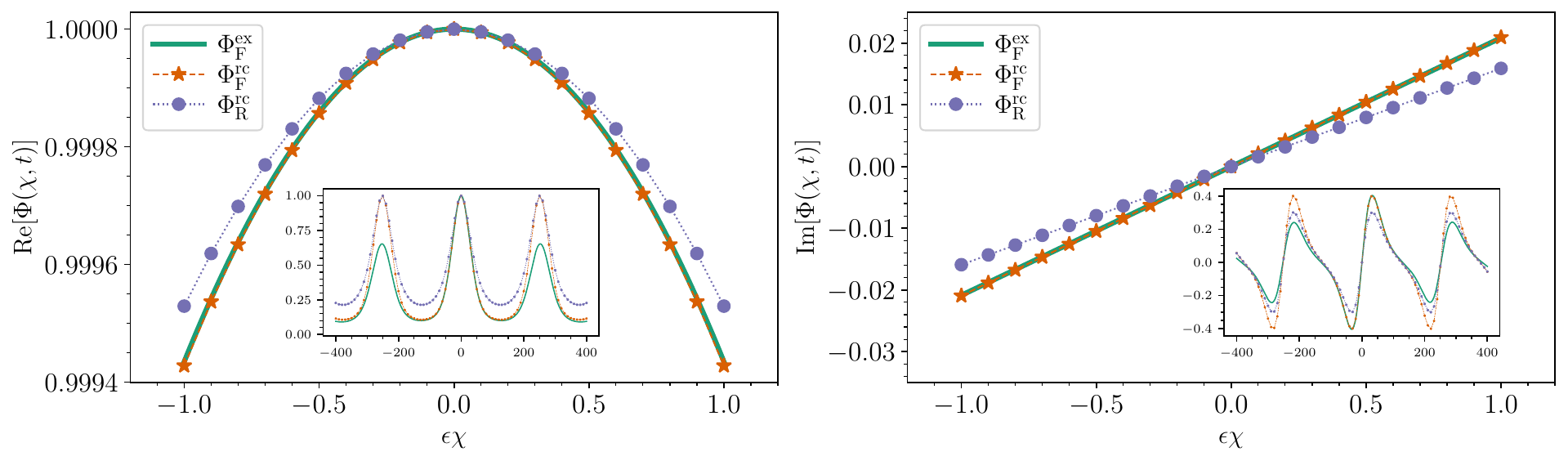}
        \caption{Real (left) and imaginary (right) parts of the CFs for different definitions of heat  transfer in the IBM, calculated with different methods. We compare an exact analytic solution $\Phi_{\mathrm{F}}^{\mathrm{ex}}$ (green, solid) with the approximate CF which uses the HC-RCME method $\Phi_{\mathrm{F}}^{\mathrm{rc}}$ (orange stars, dashed) both of which define heat as changes in the energy of the full environment. Together with this we show the CF which defines heat as energetic changes in the Hamiltonian of the residual environment, $\Phi_{\mathrm{R}}^{\mathrm{rc}}$ (purple circles, dotted), which also relies on the HC-RCME. The CFs are evaluated at a time when the TLS coherence is in the steady state. The main curves show the CFs close to $\chi=0$, where we see excellent agreement between the analytic ($\Phi_{\mathrm{F}}^{\mathrm{ex}}$) and HC-RCME ($\Phi_{\mathrm{F}}^{\mathrm{rc}}$) solutions, suggesting the HC-RCME can capture the energy statistics of the full environment. The insets show a larger range of counting parameters, where we see deviations between the analytic and HC-RCME results for the full environment. We set $\Delta=0$, the TLS energy splitting is $\epsilon=2\mathrm{eV}$, and the initial system state is $\hat{\rho}_{\mathrm{S}}(0) = \dyad{+}$. The environment spectral density parameters are $\alpha = 0.1\mathrm{eV}$, $\Gamma = 0.001\mathrm{eV}$, $\omega_{0} = 0.05\mathrm{eV}$, and the environment temperature is $T=300$K, with $N_{\mathrm{RC}} = 20$ states simulated in the RC. These parameters are used throughout the rest of this work.}
        \label{fig: cf large}
\end{figure*}

In Fig.~\ref{fig: cf large} we plot the real (left) and imaginary (right) parts of $\Phi_{\mathrm{F}}^{\mathrm{ex}}$ (green, solid) and $\Phi_{\mathrm{F}}^{\mathrm{rc}}$ (orange stars, dashed), both of which count energetic changes in the full environment Hamiltonian. We see excellent agreement between these two CFs close to $\chi=0$, with deviations at larger values of $\chi$ shown in the insets. {The analytic result shows decaying repetitions of the main feature centred around $\chi=0$, whereas the HC-RCME captures this main feature but lacks the subsequent decay for larger $\chi$ values, due to the approximations made during its derivation.} 
Also shown in Fig.~\ref{fig: cf large} is the CF which counts energetic changes in the residual environment Hamiltonian, $\Phi_{\mathrm{R}}^{\mathrm{rc}}$ (purple circles, dotted). 
We see similar qualitative behaviour between $\Phi_{\mathrm{F}}^{\mathrm{rc}}$ and $\Phi_{\mathrm{R}}^{\mathrm{rc}}$, in that they repeat their feature centered at $\chi=0$, with no decay as we increase $\chi$. 

It is clear that there are quantitative differences in the CFs when counting only on the residual environment rather than the full environment. The physical meaning behind these differences is not immediately obvious from these results alone. To gain physical insight, in the following section we calculate the first two moments of the probability distributions associated with these CFs.

\subsection{Statistical moments} 
\label{sec: moments}
The $n^{\mathrm{th}}$ moment of the probability distribution associated with a CF is given by
\begin{align}
    \label{eq: moments from derivative}
    \langle Q^{n} \rangle (t) = (-\ii)^{n}\frac{\dd^{n}}{\dd\chi^{n}}\Phi(\chi,t) \bigg|_{\chi=0}.
\end{align}
Thus, the excellent agreement between $\Phi_{\mathrm{F}}^{\mathrm{ex}}$ and $\Phi_{\mathrm{F}}^{\mathrm{rc}}$ around $\chi=0$ shown in Fig.~\ref{fig: cf large} implies that the HC-RCME can accurately capture the lower-order statistical moments of the probability distributions 
studied here~{\footnote{{We assume that if the HC-RCME captures the statistical moments of the full environment definition of heat, then it should also capture the statistical moments of the residual environment definition of heat, given that both are based on the same master equation. We note that low order moments of the heat current for both definitions could also be found directly from the RCME by tracing with the appropriate Hamiltonians.}}}.

We can find analytic forms for the moments of the full environment heat distribution by applying Eq.~\eqref{eq: moments from derivative} to Eq.~\eqref{eq: exact CF}, giving
\begin{align}
    \label{eq: exact mean}
    \langle Q_{\mathrm{F}}^{\mathrm{ex}}(t) \rangle = 2\int\limits_{0}^{\infty}\dd\omega \frac{J_{\mathrm{FE}}(\omega)}{\omega}\big(1-\cos(\omega t)\big),
\end{align}
for the mean, and 
\begin{align}
    \label{eq: exact var}
    \mathrm{var}[Q_{\mathrm{F}}^{\mathrm{ex}}(t)]  = 2\int\limits_{0}^{\infty}\dd\omega J_{\mathrm{FE}}(\omega)\coth(\frac{\beta \omega}{2})\big(1-\cos(\omega t)\big),
\end{align}
for the variance.

To calculate the mean and variance predicted by the approximate CFs, $\Phi_{\mathrm{F}}^{\mathrm{rc}}$ and $\Phi_{\mathrm{R}}^{\mathrm{rc}}$, we follow the finite-difference method used by Popovic \emph{et al}~\cite{popovic2021quantum}. By choosing a small value of the counting parameter $\chi_{\delta}$, we find the mean as
\begin{align}
    \label{eq: RC mean}
    \langle Q_{b}^{\mathrm{rc}}(t) \rangle &= \frac{\mathrm{Im}[\Phi_{b}^{\mathrm{rc}}(\chi_{\delta}, t)]}{\chi_{\delta}} + \mathcal{O}(\chi_{\delta}),
\end{align}
and the variance as
\begin{align}
    \label{eq: RC var}
    \mathrm{var}[Q_{b}^{\mathrm{rc}}(t)] &=  \frac{2 - 2\mathrm{Re}[\Phi_{b}^{\mathrm{rc}}(\chi_{\delta}, t)]}{\chi_{\delta}^{2}}-
     \langle Q_{b}^{\mathrm{rc}}(t) \rangle^2 +\mathcal{O}(\sqrt{2}\chi_{\delta}),
\end{align}
where $b = \mathrm{F}$ for the full environment definition and $b=\mathrm{R}$ for the residual environment definition. In the above $\mathcal{O}(x)$ represents error of order $x$.

\begin{figure*}[t]
\includegraphics[width=0.995\textwidth]{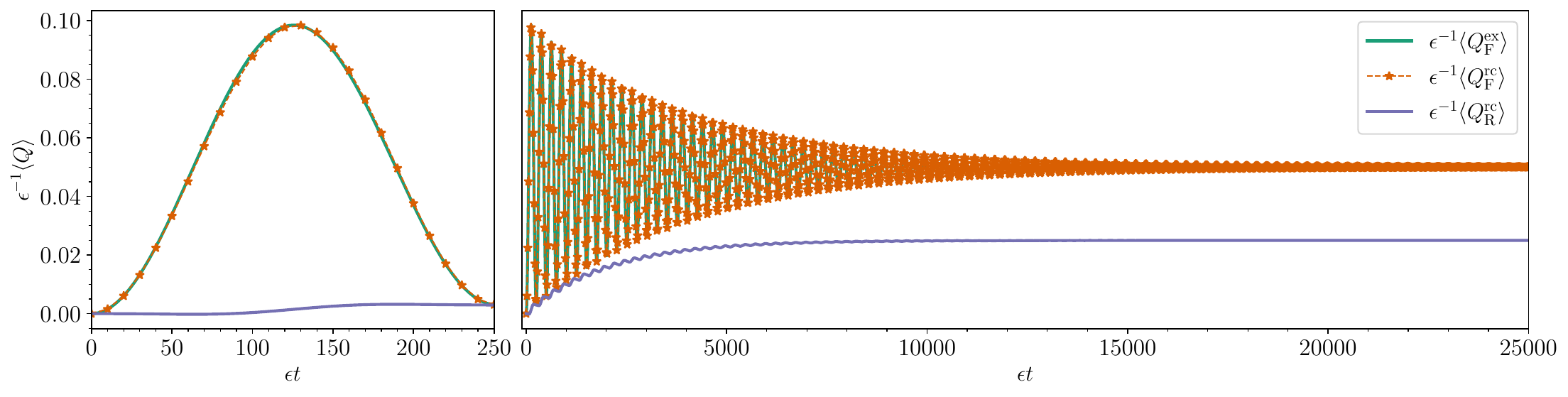}
         \includegraphics[width=0.995\textwidth]{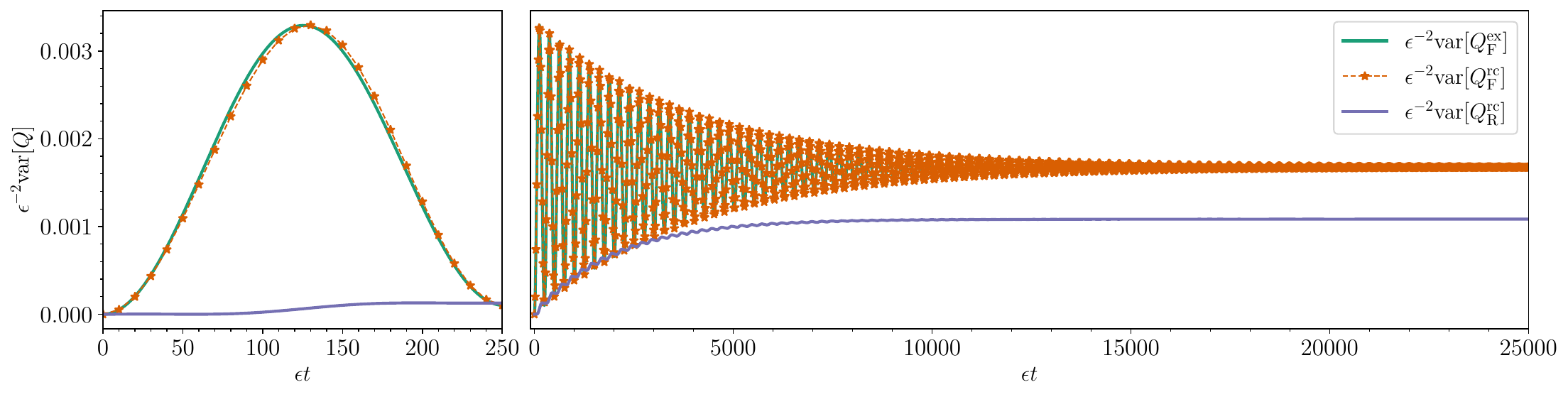}
        \caption{Mean (top) and variance (bottom) of heat transfer in the IBM, as defined by changes in the energy of the full environment, calculated using the analytic result $Q_{\mathrm{F}}^{\mathrm{ex}}$ (green, solid) and using the HC-RCME $Q_{\mathrm{F}}^{\mathrm{rc}}$ (orange stars, dashed), and changes in the energy of the residual environment $Q_{\mathrm{R}}^{\mathrm{rc}}$ using the HC-RCME (purple, solid). We show both short time (left) and long time (right) behaviour of the first two statistical moments, where we see significant qualitative differences in the mean and variance of energy changes of the full environment and the residual environment. The small value of the counting parameter we use to calculate the moments in Eq.\eqref{eq: RC mean} and Eq.\eqref{eq: RC var} is $\chi_{\delta} = 0.01/\epsilon$, and all other parameters are the same as those given in Fig.~\ref{fig: cf large}.} 
        \label{fig: moments}
\end{figure*}

In Fig.~\ref{fig: moments}~(top) we plot the mean energy change of the full environment, calculated using the exact method $\langle Q_{\mathrm{F}}^{\mathrm{ex}}\rangle$ (green, solid) and the HC-RCME method $\langle Q_{\mathrm{F}}^{\mathrm{rc}}\rangle$ (orange stars, dashed), as well as the mean energy change of the residual environment  $\langle Q_{\mathrm{R}}^{\mathrm{rc}}\rangle$ (purple, solid), showing both short (left) and long (right) timescales. 
We clearly see that the HC-RCME accurately describes the mean energy change of the full environment.
Interestingly, this definition of heat has large oscillations present, which can be intuitively explained using the RC mapping. The TLS strongly couples to the RC, which captures the long memory effects of the full environment. This leads to a coherent exchange of energy and information between the TLS and RC, suggesting that `heat', as defined by the change in energy of the full environment, is not irreversibly lost to the environment, but can re-excite~\footnote{In the independent boson model that we test, the full environment cannot excite the TLS due to the TLS and interaction Hamiltonians commuting. The change in energy of the full environment is accounted for by an equal and opposite change in the interaction Hamiltonian.} the TLS leading to coherent oscillations.

This contrasts with the mean heat predicted by counting on only the residual environment, which contains heavily suppressed oscillations, suggesting that energy and information that leave the extended system is lost irreversibly to the residual environment. 
This behaviour is more in keeping with the classical definition of heat, which is understood to be the changes in internal energy which lead to a monotonic entropy change, and which flows in a unidirectional manner from a hot body to a cold body.  


These differences are also reflected in Fig.~\ref{fig: moments}~(bottom), where we plot the variances associated with the CFs. Once again, we see excellent agreement between the exact treatment and the HC-RCME when counting on the full environment. 
Similar to the case of the mean, we see both quantitative and qualitative differences in the variance when counting only on the residual environment, with coherent oscillations suppressed. The variance of heat transfer is expected to monotonically increase with time, as seen in the residual environment definition, suggesting it is most suitable for the non-Markovian regime.

\subsection{Changes in ergotropy and entropy} \label{sec: ergotropy}
\begin{figure*}[t]
     \centering
        \includegraphics[width=0.995\textwidth]{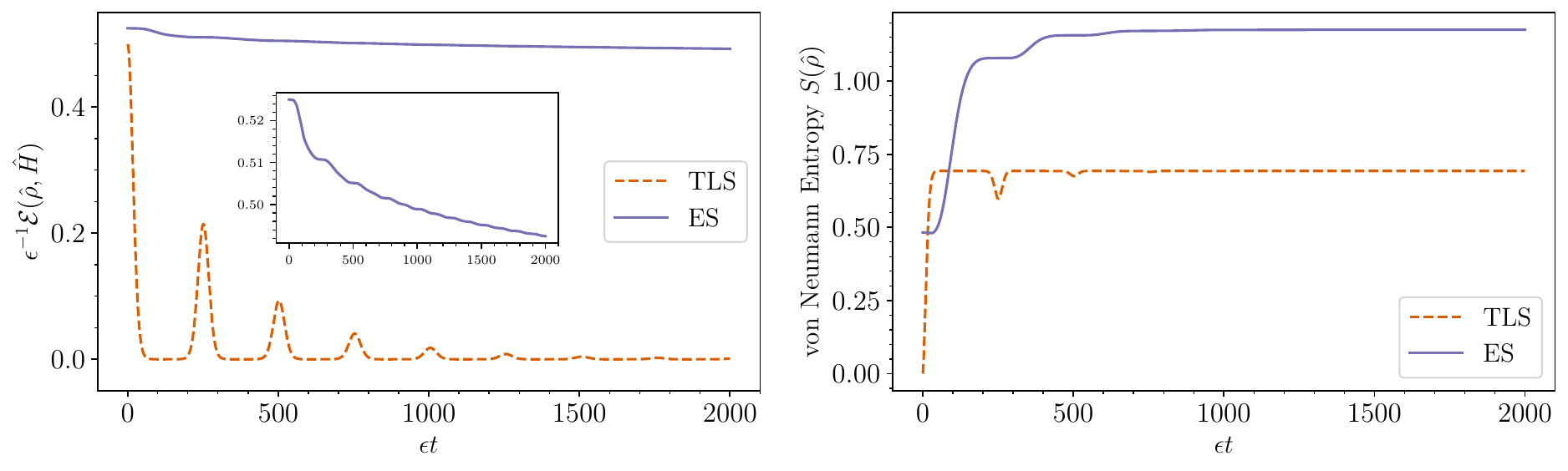}
        \caption{Left: Ergotropy of the TLS (orange dashed) and extended system (purple, solid). Decay and periodic revivals of the TLS ergotropy are due to decoherence and re-coherence of the TLS, respectively. The ergotropy of the extended system remains higher than that of the TLS at all times, as the RC provides access to coherence that is otherwise inaccessible in the TLS alone. Inset: Decay of the extended system ergotropy is due to its weak interaction with the residual environment, which causes irreversible decoherence and dissipation. Right: von Neumann entropy of the TLS (orange dashed) and extended system (purple, solid). Decoherence and re-coherence of the TLS leads to an entropy which does not change monotonically, indicating that the TLS is not in contact with a simple heat bath. All physical parameters are the same as those given in Fig.~\ref{fig: cf large}.}
        \label{fig: ergotropy}
\end{figure*}
The coherent oscillations observed in Fig.~\ref{fig: moments} for both methods of calculating the mean energy change of the full environment, $\langle Q_{\mathrm{F}}^{\mathrm{ex}}\rangle$ and $\langle Q_{\mathrm{F}}^{\mathrm{rc}}\rangle$, suggests that there is a work-like contribution within this definition of heat.
The RC method provides us with a unique insight into this behaviour, providing an avenue to calculate the ergotropy ~\cite{allahverdyan2004maximal, perarnau2015extractable, sone2021quantum} of the TLS, which then treats all energy emitted into the environment as heat (i.e. the full environment paradigm referenced above), as well as the ergotropy of the extended system, which treats the RC as a potential work source. 

Ergotropy is defined as the maximum amount of work we can extract unitarily from a quantum state, and for the two environment partitions can be defined as, 
\begin{align}
\label{eq: sys ergo}
    \mathcal{E}(\hat{\rho}_{\mathrm{S}}, \hat{H}_{\mathrm{S}}) = \Tr[\hat{H}_{\mathrm{S}}\hat{\rho}_{\mathrm{S}}] - \min_{\hat{U}} \Tr[\hat{H}_{\mathrm{S}} \hat{U} \hat{\rho}_{\mathrm{S}} \hat{U}^{\dagger}],
\end{align}
for the TLS, and 
\begin{align}
\label{eq: ext sys ergo}
    \mathcal{E}(\hat{\rho}_{\mathrm{ES}}, \hat{H}_{\mathrm{ES}}) = \Tr[\hat{H}_{\mathrm{ES}}\hat{\rho}_{\mathrm{ES}}] - \min_{\hat{V}} \Tr[\hat{H}_{\mathrm{ES}} \hat{V} \hat{\rho}_{\mathrm{ES}} \hat{V}^{\dagger}],
\end{align}
for the extended system.
The operators $\hat{U}$ and $\hat{V}$ act on the Hilbert spaces of the TLS and extended system, respectively~\footnote{The time dependence of the ergotropies and states have been omitted from the above definition for ease of reading.}. The minimisation is satisfied through a unitary transformation which takes the state to its passive counterpart, from which no further work can be extracted~\cite{allahverdyan2004maximal}. By diagonalising the state as
\begin{align}
\label{eq: rho diagonlised}
    \hat{\rho} = \sum_{j}r_{j}\dyad{r_{j}}{r_{j}},
\end{align}
such that $r_{j} > r_{j+1}$, and diagonalising the Hamiltonian as
\begin{align}
    \label{eq: Hamiltonian diagonlised}
    \hat{H} = \sum_{k}h_{k}\dyad{h_{k}}{h_{k}},
\end{align}
such that $h_{k} < h_{k+1}$ the ergotropy is given by
\begin{align}
    \label{eq: ergotropy diagonalised}
    \mathcal{E}(\hat{\rho}, \hat{H}) =\sum_{jk}r_{j}h_{k}\Big(\abs{\bra{r_{j}}\ket{h_{k}}}^{2} - \delta_{jk}\Big).
\end{align}

In Fig.~\ref{fig: ergotropy} (left) we plot the ergotropies of the TLS state and extended system state, when initialising the TLS in $\hat{\rho}_{\mathrm{S}}(0) = \dyad{+}$ and the RC in the Gibbs state. At $t=0$ the TLS ergotropy (orange, dashed) is the amount of extractable energy by unitarily transforming the state from $\ket{+}$ to the ground state $\ket{g}$, as expected. 
We then see a sharp decrease in ergotropy followed by decaying recurrences, which are in phase with the decoherence and re-coherences of the TLS induced by the non-Markovian interaction with the full environment, as seen in Fig.~\ref{fig:sx dynamics}. This is explained by coherence in the energy eigenbasis contributing to ergotropy~\cite{francica2020quantum, ccakmak2020ergotropy, touil2021ergotropy}.
When calculating the erogtropy of the TLS state, we assume we have complete control over the TLS, but no control over the full environment. During decoherence, the TLS cannot use coherence in the energy eigenbasis as a source of extractable work, and hence the ergotropy decreases. During the re-coherence process, the full environment generates coherence in the TLS (in ever smaller quantities), providing an increase in the extractable work (also in ever smaller quantities). 

When considering the ergotropy of the extended system, we 
see that the presence of the RC increases ergotropy by providing access to coherence that are otherwise lost to the full environment.
Notably, the ergotropy of the extended system begins at a larger value than that of the TLS, despite the RC beginning in the Gibbs state. While the Gibbs state of the RC is passive with respect to the RC Hamiltonian, the presence of the TLS-RC interaction Hamiltonian means the Gibbs state is not passive with respect to the extended system Hamiltonian, thus providing an increase in the amount of extractable work for the extended system state. This interaction maintains coherence within the extended system, explaining the (relatively) constant ergotropy. However, a gradual decay occurs in the extended system's ergotropy resulting from the weak interaction between the RC and the (infinite) residual environment, causing slow, and irreversible decoherence and dissipation in the extended system, as seen in the inset of Fig.~\ref{fig: ergotropy} (left).


Next, we calculate the von Neumann entropy for the TLS and the extended system states. In the IBM, if we begin the TLS in a pure state (i.e. $\ket{+}$), we know it will dephase due to its interaction with the full environment. We therefore expect the entropy of the TLS to increase as the system loses coherence and becomes more mixed, and to decrease during any re-coherence process as it becomes purer. However, were the system to interact with a heat sink, 
its entropy should change monotonically. 

In Fig.~\ref{fig: ergotropy} (right), we see that the von Neumann entropy of the TLS (orange dashed) does not change monotonically, which adds evidence to the argument that the TLS is not in contact with a heat bath (i.e. changes in the full environment cannot be characterised solely as heat in this regime). Notice that the changes in entropy of the TLS are in phase with changes in its ergotropy, both of which are in phase with the coherence of the TLS shown in Fig.~\ref{fig:sx dynamics}. As the TLS regains coherence via a non-Markovian interaction, its state becomes purer, reducing entropy and increasing ergotropy. Meanwhile, the extended system’s entropy increases monotonically~\footnote{Except for a slight decrease within the first few time steps.}, adding further evidence to the fact that the residual environment acts as a heat bath.

These results appear to show that the RC is a viable work source, at least in principle~\footnote{We are not suggesting a method of extracting work from the RC, but are using the fact that work can in principle be extracted from it to make arguments about the most appropriate definition of heat.}. Therefore, if work and heat are to be defined as separate and distinct components of a system's internal energy change, definitions of heat that do not include energetic changes within the RC may be more appropriate than those that do. This gives further evidence to support using the definition of heat as energetic changes in the residual environment Hamiltonian when studying the non-Markovian regime.

\section{Discussion} \label{sec: conc}
We have used the two-point measurement protocol to study the full counting statistics of energetic changes in the independent boson model within the non-Markovian regime. We employed the reaction coordinate formalism,  
which not only accounts for TLS-environment correlations through the RC - enabling us to study energetic changes in the environment in strong coupling and non-Markovian regimes - but also allows us to investigate how different system-environment partitions impact the definition of heat in open quantum systems. {In future work it would be interesting to study the similarities and differences between secular and non-secular versions of the HC-RCME, as well as the behaviour of the heat definition when performing the reaction coordinate mapping multiple times.}

Notably, we find that a na\"ive definition of heat, in which all energy changes of an environment is categorised as heat, overlooks significant work-like contributions present in the TLS-environment interactions, which leads to coherent oscillations in both the average and variance of these energetic changes.
The RC formalism, however, allows us to use an alternative definition in which only energetic changes of the residual environment are considered as heat.
With this definition, coherent oscillations in the mean and variance of heat are heavily suppressed, suggesting that heat transfer within the independent boson model in this case is truly an irreversible process. 
This is supported by considering the ergotropy and von Neumann entropy for both the TLS and extended system.

This work demonstrates that defining heat as changes in the energy of the residual environment, as characterized by the RC mapping, aligns with the classical intuition of heat: monotonic entropy non-conserving changes in internal energy that are distinct from work transfer. Recent research by Colla \emph{et al.}~\cite{colla2024thermodynamic} supports this conclusion, revealing that peaks in the spectral density of a supposed heat bath can enable it to partially function as a work reservoir. Combined with our findings, this suggests that assumptions about the interaction between an open quantum system and a thermal environment must be made carefully when considering quantum thermodynamics. It is thus particularly crucial to consider the structure of the environment's spectral density when evaluating energy transfer and distinguishing between heat and work.

\section{Acknowledgements}
MS acknowledges support from UK EPSRC (EP/SO23607/1) and (EP/W524347/1). We thank Harry Miller for discussions.
\onecolumngrid
\begin{appendix}

\section{Benchmarking dynamics}
\label{app:bench}
\begin{figure*}[t]
     \centering
     \includegraphics[width=0.995\textwidth]{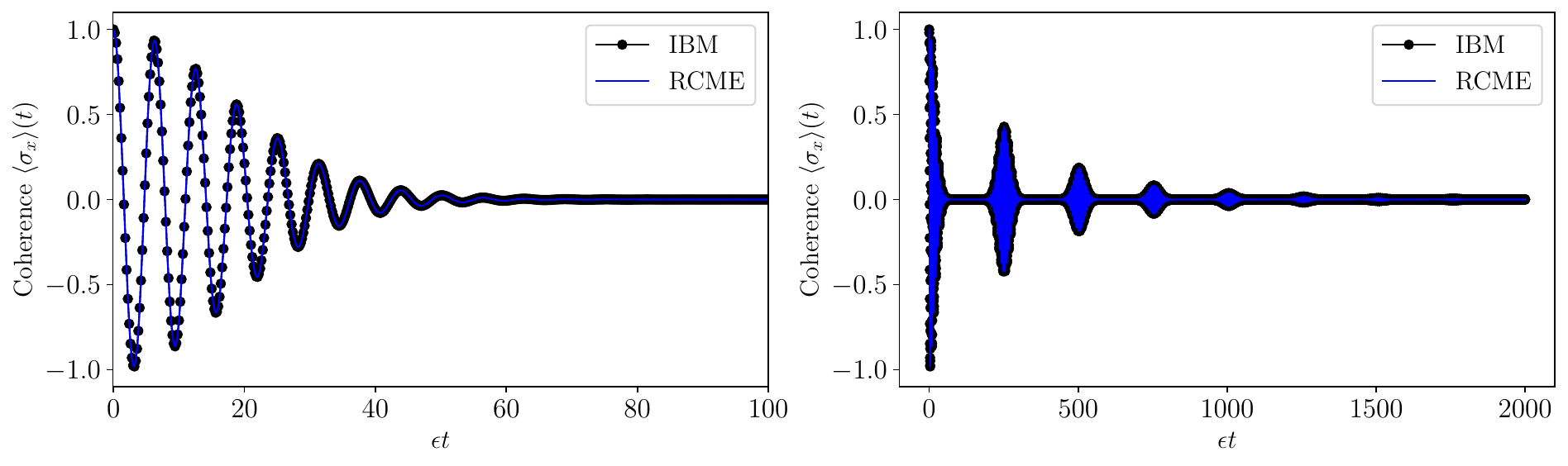}
        \caption{Coherence dynamics $\langle \hat{\sigma}_{x}(t)\rangle = \Tr[\hat{\sigma}_{x}\hat{\rho}_{\mathrm{S}}(t)]$ for the TLS in the independent boson model, using the exact analytic result (black dots) and the RCME (blue) for both a short (left) and long (right) timescale. We take $N_{\mathrm{RC}}=20$ energy levels in the RC. We set $\Delta=0$, the TLS energy splitting is $\epsilon=2\mathrm{eV}$, and the initial system state is $\hat{\rho}_{\mathrm{S}}(0) = \dyad{+}$. The environment spectral density parameters are $\alpha = 0.1\mathrm{eV}$, $\Gamma = 0.001\mathrm{eV}$, $\omega_{0} = 0.05\mathrm{eV}$, and the environment temperature is $T=300$K.}
        \label{fig:sx dynamics}
\end{figure*}

Here we show that the RCME (HC-RCME with $\chi=0)$ is able to track TLS properties accurately. 
In Fig.~\ref{fig:sx dynamics} we plot the dynamics of the TLS coherence $\langle \hat{\sigma}_{x}(t)\rangle$ on both a short (left) and long (right) timescale, by solving the RCME (blue) and from the analytic result for the IBM (black dots). We see that the RCME is able to track the TLS coherence dynamics very well with $N_{\mathrm{RC}}=20$ energy levels being simulated in the RC. The TLS energy splitting is $\epsilon=2\mathrm{eV}$, the environment spectral density parameters are $\alpha = 0.1\mathrm{eV}$, $\Gamma = 0.001\mathrm{eV}$, $\omega_{0} = 0.05\mathrm{eV}$, and the environment temperature is $T=300$K (as elsewhere). 
The TLS coherence shows decaying oscillations on a short timescale, Fig.~\ref{fig:sx dynamics} (left), with periodic re-coherences on longer timescales, which themselves decay in time, Fig.~\ref{fig:sx dynamics} (right). These re-coherences are a result of the sharply peaked form of the full environment spectral density, and indicate that we are working in a regime where the standard Born-Markov master equation would fail, as it is unable to predict re-coherences of the TLS.

\section{Generalised master equation in the reaction coordinate frame}
\label{app: HC-RCME}
The interaction Hamiltonian in Eq.~\eqref{eq:RC_INT} can be written as $\hat{H}_{\mathrm{I}} = \hat{A}\otimes \hat{B}$, where $\hat{A}$ and $\hat{B}$ are operators on the Hilbert spaces of the extended system and residual environment, respectively. We ignore the counter term since it does not contribute to the master equation we are about to derive~\cite{iles2014environmental,nazir2018reaction}.
Dressing this interaction Hamiltonian with the counting parameter and moving into the interaction picture with respect to $\hat{H}_{\mathrm{ES}} + \hat{H}_{\mathrm{RB}}$ gives
\begin{align}
     \tilde{H}_\mathrm{I}(\chi,t) &= \mathrm{e}^{\ii \hat{H}_{\mathrm{ES}}t}\hat{A}\mathrm{e}^{-\ii \hat{H}_{\mathrm{ES}}t}
     \otimes \mathrm{e}^{\ii \hat{H}_{\mathrm{RE}}t}\mathrm{e}^{\ii\frac{\chi}{2}\hat{H}_{\mathrm{RE}}} \hat{B}\mathrm{e}^{-\ii\frac{\chi}{2}\hat{H}_{\mathrm{RE}}}\mathrm{e}^{-\ii \hat{H}_{\mathrm{RE}}t}= \Tilde{A}(t)\otimes \Tilde{B}(\chi,t).
\end{align}

Upon applying the Born-Markov approximations between the extended system and residual environment we arrive at the `heat-counting reaction coordinate master equation' (HC-RCME)
\begin{align}
\label{ch:heat, eq: CFDRCME}
    \dt \hat{\rho}_{\mathrm{ES}}(\chi, t) &= -\ii\comm{\hat{H}_{\mathrm{ES}}}{\hat{\rho}_{\mathrm{ES}}(\chi, t)} - \hat{A} \hat{A}_{1}\hat{\rho}_{\mathrm{ES}}(\chi, t)  + \hat{A}\hat{\rho}_{\mathrm{ES}}(\chi, t) \hat{A}_{2}(\chi) + \hat{A}_{3}(\chi)\hat{\rho}_{\mathrm{ES}}(\chi, t) \hat{A} - \hat{\rho}_{\mathrm{ES}}(\chi, t)\hat{A}_{4}\hat{A} \nt&=  \mathcal{L}_{\mathrm{ES}}(\chi)[\hat{\rho}_{\mathrm{ES}}(\chi, t)].
\end{align}
The terms in the above equation are given by
\begin{align}
\label{eq: rate operator 1}
    \hat{A}_{1} & = \int\limits_{0}^{\infty}\dd\tau\tilde{A}(-\tau)\langle\tilde{B}(\chi,\tau)\hat{B}(\chi)\rangle,\\
    \label{eq: rate operator 2}
    \hat{A}_{2}(\chi) & = \int\limits_{0}^{\infty}\dd\tau\tilde{A}(-\tau)\langle\tilde{B}(-\chi,-\tau)\hat{B}(\chi)\rangle,\\
    \label{eq: rate operator 3}
    \hat{A}_{3}(\chi) & = \int\limits_{0}^{\infty}\dd\tau\tilde{A}(-\tau)\langle\tilde{B}(-\chi,\tau)\hat{B}(\chi)\rangle,\\
    \label{eq: rate operator 4}
    \hat{A}_{4} & = \int\limits_{0}^{\infty}\dd\tau\tilde{A}(-\tau)\langle\tilde{B}(-\chi,-\tau)\hat{B}(-\chi)\rangle,
\end{align}
with
\begin{align}
    \label{ch:heat, eq: A(-tau)}
    \tilde{A}(-\tau) &= \sum_{j,k=1}^{2N_{\mathrm{RC}}}\mathrm{e}^{-\ii\lambda_{jk}\tau} A_{jk}\dyad{\lambda_{j}}{\lambda_{k}},
\end{align}
where the extended system Hamiltonian in its spectral form is given by $\hat{H}_{\mathrm{ES}} = \sum_{k=1}^{2N_{\mathrm{RC}}}\lambda_{k}\dyad{\lambda_{k}}{\lambda_{k}}$, and where the $\chi$-dependent residual environment correlation functions are given by
\begin{align}
    \label{ch:heat, heat counting, environment correlation functions}
    \langle\tilde{B}(\chi, \tau)\hat{B}(\chi)\rangle &= \int\limits_{0}^{\infty}\dd \omega J_{\mathrm{RE}}(\omega)\bigg(N(\omega)\mathrm{e}^{\ii\omega \tau} + (1+N(\omega))\mathrm{e}^{-\ii\omega \tau})\bigg),
    \end{align} 
\begin{align}
    \langle\tilde{B}(-\chi, \pm\tau)\hat{B}(\chi)\rangle &= \int\limits_{0}^{\infty}\dd \omega J_{\mathrm{RE}}(\omega)\bigg(N(\omega)\mathrm{e}^{\pm\ii\omega \tau}\mathrm{e}^{-\ii\chi\omega} + (1+N(\omega))\mathrm{e}^{\mp\ii\omega \tau}\mathrm{e}^{\ii\chi\omega}\bigg),
    \end{align} 
\begin{align}
    \langle\tilde{B}(-\chi, -\tau)\hat{B}(-\chi)\rangle &= \int\limits_{0}^{\infty}\dd \omega J_{\mathrm{RE}}(\omega)\bigg(N(\omega)\mathrm{e}^{-\ii\omega \tau} + (1+N(\omega))\mathrm{e}^{\ii\omega \tau})\bigg).
\end{align}
In the above, $N(\omega) = (\mathrm{e}^{\beta \omega}-1)^{-1}$ is the thermal occupation number for bosons. By making use of the Sokhotski-Plemelj theorem~\cite{breuer2002theory}
\begin{align}
    \int\limits_{0}^{\infty}\dd\tau \mathrm{e}^{-\ii (\omega \pm \nu)\tau} = \pi\delta(\omega \pm \nu) - \ii \mathcal{P}\bigg(\frac{1}{\omega \pm \nu}\bigg),
\end{align}
we can evaluate the time and frequency integrals. In the above, $\delta(x-a)$ represents a Dirac delta function centred at $x=a$, and $\mathcal{P}$ is the Cauchy principal value. 
After evaluating the 
integrals, ignoring the principal value terms, and taking into consideration the three possible cases for $\lambda_{mn}$, we obtain 
\begin{align}
\label{ch:heat, eq: A1}
    \hat{A}_{1}
    &= \sum_{mn}\begin{cases}
         \pi A_{mn}J_{\mathrm{RE}}(\lambda_{mn})N(\lambda_{mn})\dyad{\lambda_{m}}{\lambda_{n}}& \mathrm{if } \lambda_{mn} > 0 , \\
          \pi A_{mn}J_{\mathrm{RE}}(\abs{\lambda_{mn}})(1+N(\abs{\lambda_{mn}}))\dyad{\lambda_{m}}{\lambda_{n}} & \mathrm{if } \lambda_{mn} < 0, \\
        \pi A_{mn}\gamma\beta^{-1}\dyad{\lambda_{m}}{\lambda_{n}}  & \mathrm{if } \lambda_{mn} = 0,
\end{cases}
\end{align} 
\begin{align}
\label{ch:heat, eq: A2}
    \hat{A}_{2}(\chi)
    &= \sum_{mn}\begin{cases}
         A_{mn}J_{\mathrm{RE}}(\lambda_{mn})(1+N(\lambda_{mn}))\mathrm{e}^{\ii\chi \lambda_{mn}}\dyad{\lambda_{m}}{\lambda_{n}}& \mathrm{if } \lambda_{mn} > 0 , \\
          \pi A_{mn}J_{\mathrm{RE}}(\abs{\lambda_{mn}})N(\abs{\lambda_{mn}})\mathrm{e}^{-\ii\chi \abs{\lambda_{mn}}}\dyad{\lambda_{m}}{\lambda_{n}} & \mathrm{if } \lambda_{mn} < 0, \\
        \pi A_{mn}\gamma\beta^{-1}\dyad{\lambda_{m}}{\lambda_{n}}  & \mathrm{if } \lambda_{mn} = 0,
\end{cases}
\end{align} 
\begin{align}
\label{ch:heat, eq: A3}
    \hat{A}_{3}(\chi)
    &= \sum_{mn}\begin{cases}
         \pi A_{mn}J_{\mathrm{RE}}(\lambda_{mn})N(\lambda_{mn})\mathrm{e}^{-\ii\chi \lambda_{mn}}\dyad{\lambda_{m}}{\lambda_{n}}& \mathrm{if } \lambda_{mn} > 0 , \\
          \pi A_{mn}J_{\mathrm{RE}}(\abs{\lambda_{mn}})(1+N(\abs{\lambda_{mn}}))\mathrm{e}^{\ii\chi \abs{\lambda_{mn}}}\dyad{\lambda_{m}}{\lambda_{n}} & \mathrm{if } \lambda_{mn} < 0, \\
        \pi A_{mn}\gamma\beta^{-1}\dyad{\lambda_{m}}{\lambda_{n}}  & \mathrm{if } \lambda_{mn} = 0,
\end{cases}
\end{align} 
\begin{align}
\label{ch:heat, eq: A4}
    \hat{A}_{4}
    &= \sum_{mn}\begin{cases}
         \pi A_{mn}J_{\mathrm{RE}}(\lambda_{mn})(1+N(\lambda_{mn}))\dyad{\lambda_{m}}{\lambda_{n}}& \mathrm{if } \lambda_{mn} > 0 , \\
          \pi A_{mn}J_{\mathrm{RE}}(\abs{\lambda_{mn}})N(\abs{\lambda_{mn}})\dyad{\lambda_{m}}{\lambda_{n}} & \mathrm{if } \lambda_{mn} < 0, \\
        \pi A_{mn}\gamma\beta^{-1}\dyad{\lambda_{m}}{\lambda_{n}}  & \mathrm{if } \lambda_{mn} = 0.
\end{cases}
\end{align} 
If we set $\chi=0$ we recover the standard reaction coordinate master equation.

\end{appendix}

\bibliographystyle{unsrt}
\bibliography{bib.bib}

\end{document}